\documentclass[12pt]{iopart}
\usepackage{latexsym}
\usepackage{amscd,amsfonts}
\usepackage[dvips]{graphics,graphics,color}
\begin{document}
\definecolor{r}{rgb}{1,0,0}
\definecolor{b}{rgb}{0,0,1}
\definecolor{g}{cmyk}{0,0,1,0}
\jl{1}

\jl{1}
 \def\lambdabar{\protect\@lambdabar}
\def\@lambdabar{%
\relax
\bgroup
\def\@tempa{\hbox{\raise.73\ht0
\hbox to0pt{\kern.25\wd0\vrule width.5\wd0
height.1pt depth.1pt\hss}\box0}}%
\mathchoice{\setbox0\hbox{$\displaystyle\lambda$}\@tempa}%
{\setbox0\hbox{$\textstyle\lambda$}\@tempa}%
{\setbox0\hbox{$\scriptstyle\lambda$}\@tempa}%
{\setbox0\hbox{$\scriptscriptstyle\lambda$}\@tempa}%
\egroup
}

\def\bbox#1{%
\relax\ifmmode
\mathchoice
{{\hbox{\boldmath$\displaystyle#1$}}}%
{{\hbox{\boldmath$\textstyle#1$}}}%
{{\hbox{\boldmath$\scriptstyle#1$}}}%
{{\hbox{\boldmath$\scriptscriptstyle#1$}}}%
\else
\mbox{#1}%
\fi
}
\newcommand{\muv}{\bbox{\mu}}
\newcommand{\mc}{{\mathcal M}}
\newcommand{\pc}{{\mathcal P}}
\newcommand{\mct}{\bbox{\mathcal M}}
\newcommand{\pct}{\bbox {\mathcal P}}
\newcommand{\fsf}{{\sf F}}
\newcommand{\fsft}{\bbox{{\sf F}}}
 \newcommand{\mv}{\bbox{m}}
\newcommand{\pv}{\bbox{p}}
\newcommand{\tv}{\bbox{t}}
\def\msf{\hbox{{\sf M}}}
\def\msft{\bbox{{\sf M}}}
\def\psf{\hbox{{\sf P}}}
\def\psft{\bbox{{\sf P}}}
\def\Nsf{\hbox{{\sf N}}}
\def\Nsft{\bbox{{\sf N}}}
\def\Tsf{{\sf T}}
\def\Tsft{\bbox{{\sf T}}}
\def\Asf{\hbox{{\sf A}}}
\def\Asft{\bbox{{\sf A}}}
\def\Bsf{\hbox{{\sf B}}}
\def\Bsft{\bbox{{\sf B}}}
\def\Lsf{\hbox{{\sf L}}}
\def\Lsft{\bbox{{\sf L}}}
\def\Ssf{\hbox{{\sf S}}}
\def\Ssft{\bbox{{\sf S}}}
\def\Mtens{\bi{M}}
\def\msfsim{\bbox{{\sf M}}_{\scriptstyle\rm(sym)}}
\newcommand{\mcsim}{ {\sf M}_{ {\scriptstyle \rm {(sym)} } i_1\dots i_n}}
\newcommand{\mcs}{ {\sf M}_{ {\scriptstyle \rm {(sym)} } i_1i_2i_3}}

\newcommand{\beqan}{\begin{eqnarray*}}
\newcommand{\eeqan}{\end{eqnarray*}}
\newcommand{\beqa}{\begin{eqnarray}}
\newcommand{\eeqa}{\end{eqnarray}}

 \newcommand{\suml}{\sum\limits}
 \newcommand{\sumd}{\suml_{\mathcal D}}
\newcommand{\intl}{\int\limits}
\newcommand{\rvec}{\bbox{r}}
\newcommand{\xivec}{\bbox{\xi}}
\newcommand{\Avec}{\bbox{A}}
\newcommand{\Rvec}{\bbox{R}}
\newcommand{\Evec}{\bbox{E}}
\newcommand{\Bvec}{\bbox{B}}
\newcommand{\Svec}{\bbox{S}}
\newcommand{\avec}{\bbox{a}}
\newcommand{\nablav}{\bbox{\nabla}}
\newcommand{\nuvec}{\bbox{\nu}}
\newcommand{\bvec}{\bbox{\beta}}
\newcommand{\vvec}{\bbox{v}}
\newcommand{\jvec}{\bbox{J}}
\newcommand{\nvec}{\bbox{n}}
\newcommand{\pvec}{\bbox{p}}
\newcommand{\mvec}{\bbox{m}}
\newcommand{\evec}{\bbox{e}}
\newcommand{\eps}{\varepsilon}
\newcommand{\la}{\lambda}
\newcommand{\rad}{\mbox{\footnotesize rad}}
\newcommand{\scr}{\scriptstyle}
\newcommand{\latens}{\bbox{\sf{\Lambda}}}
\newcommand{\lasf}{{\sf{\Lambda}}}
\newcommand{\pitens}{\sf{\Pi}}
\newcommand{\cm}{{\cal M}}
\newcommand{\cp}{{\cal P}}
\newcommand{\beq}{\begin{equation}} 
\newcommand{\eeq}{\end{equation}}
\newcommand{\ptens}{\bbox{\sf{P}}}
\newcommand{\Ptens}{\bbox{P}}
\newcommand{\Ttens}{\bbox{\sf{T}}}
\newcommand{\Ntens}{\bbox{\sf{N}}}
\newcommand{\Ncal}{\bbox{{\cal N}}}
\newcommand{\Atens}{\bbox{\sf{A}}}
\newcommand{\Btens}{\bbox{\sf{B}}}
\newcommand{\dom}{\mathcal{D}}
\newcommand{\al}{\alpha}
\newcommand{\sym}{\scriptstyle \rm{(sym)}}
\newcommand{\Tcal}{\bbox{{\mathcal T}}}
\newcommand{\Nmc}{{\mathcal N}}
\renewcommand{\d}{\partial}
\def\rmi{{\rm i}}
\def\rme{\hbox{\rm e}}
\def\rmd{\hbox{\rm d}}
\newcommand{\ct}{\mbox{\Huge{.}}}
\newcommand{\Laop}{\bbox{\Lambda}}
\newcommand{\Ssfs}{{\scriptstyle \Ssft^{(n)}}}
\newcommand{\Lsfs}{{\scriptstyle \Lsft^{(n)}}}
\newcommand{\psfr}{\widetilde{\psf}}
\newcommand{\msfr}{\widetilde{\msf}}
\newcommand{\msftr}{\widetilde{\msft}}
\newcommand{\psftr}{\widetilde{\psft}}
\newcommand{\pvr}{\widetilde{\pvec}}
\newcommand{\mvr}{\widetilde{\mvec}}
\newcommand{\qdot}{\stackrel{\cdot\cdot\cdot\cdot}}
\newcommand{\bsy}{\hbox}
\newcommand{\ointl}{\oint\limits}
\newcommand{\pisf}{{\sf \Pi}}
\newcommand{\ssft}{\bbox{{\sf S}}}
\newcommand{\ssf}{{\sf S}}
\def\Nsf{{\sf N}}
\newcommand{\gamsf}{{\sf \Gamma}}
\newcommand{\gamsft}{\bsy{\sf \Gamma}}
\newcommand{\mlrt}{\stackrel{\leftrightarrow}{\msft}}
\newcommand{\mlr}{\stackrel{\leftrightarrow}{\msf}}
\newcommand{\ab}{\v{a}} 
\newcommand{\ai}{\^{a}} 
\newcommand{\ib}{\^{\i}} 
\newcommand{\tb}{\c{t}} 
\newcommand{\st}{\c{s}}
\newcommand{\Ab}{\v{A}} 
\newcommand{\Ai}{\^{A}} 
\newcommand{\Ib}{\^{I}} 
\newcommand{\Tb}{\c{T}}
\newcommand{\St}{\c{S}}
\newcommand{\lavec}{\bbox{\Lambda}}
\newcommand{\esft}{\bbox{{\sf e}}}

\title{ A note on the $\delta$-singularities of the   static electric and magnetic fields  }
\author{C  Vrejoiu\footnote{E-mail:  vrejoiu@fizica.unibuc.ro}, R Zus\footnote{E-mail: roxana.zus@fizica.unibuc.ro}}
\address{University of Bucharest, Department of Physics,  \\
PO Box MG - 11, Bucharest-Magurele, RO - 077125,
 Romania }
 \begin{abstract}
The $\delta-$singularities of the electric and magnetic fields in the static case are  established based on the regularized $\delta_\eps(\rvec)$ function introduced by Jackson \cite{Jackson}. 

\end{abstract}
\section{Introduction}
The problem of  regularization  when searching the $\delta-$singularities of the multipole electromagnetic fields  is treated in the present paper in a simple procedure. In the static case, we can use a generalization of a simple identity implying  a regularized derivative of $1/r$ and the Jackson regularized function $\delta_\eps(\rvec)$ 
\cite{Jackson} and \cite{Hnizdo} - equation  (2). \\
In Section 2 we treat the electrostatic field, while in section 3, the magnetic field. In both the cases, we discuss the singularities for the first three multipoles: dipole, quadrupole and octopole. In Section 4, as a simple mathematical digression, we generalize the results for arbitrary multipole orders.\\

\section{Electrostatic field}
In the present section we search the $\delta-$singularities of the electrostatic fields of the electric dipole, quadrupole and octopole. These fields correspond to the following multipole expansion  written here in the case of an electric neutral charge distribution in a finite space region  $\dom$ \cite{Jackson}, \cite{Castell}:
\beqa\label{2.2}
\Evec(\rvec)=\frac{1}{4\pi\eps_0}\evec_i\left(p_j\d_i\d_j\frac{1}{r}-\frac{1}{2}\psf_{jk}\d_i\d_j\d_k\frac{1}{r}+
\frac{1}{6}\psf_{jkl}\d_i\d_j\d_k\d_l\frac{1}{r}+ \dots\right)\ .
\eeqa
$p_i,\,\psf_{ij},\,\psf_{ijk}$ are the Cartesian components of the electric multipole moments:
\beqan
\fl\;\;\;\;\;\;\;\;\;\;\;\;p_i=\int_{\dom}\rmd^3x\,x_i\rho(\rvec),\;\;\psf_{ij}=\int_{\dom}\rmd^3x\,x_ix_j\,\rho(\rvec),\;\;
\psf_{ijk}=\int_{\dom}\rmd^3x\,x_ix_jx_k\,\rho(\rvec)\ .
\eeqan
The origin  $O$ of the Cartesian axes is chosen in the domain $\dom$ and $\evec_i,\;i=1,2,3$ are the unit vectors of these axes. \\
The various terms from the multipole expansion \eref{2.2}, considered as the  fields corresponding to point-like multipoles,  are defined as functions on $\mathbb{R}^3 $ having as support the entire space without the point $O$. The extensions of such functions to the entire space can be realized  as distributions (generalized functions) adding 
some distributions with point-like support to the expressions defined only for $r\ne0$. These last distributions are generally linear combinations of the $\delta-$functions and their derivatives. These $\delta-$type distributions correspond, actually, to the extension of the various orders partial derivatives of $1/r$.\\
The well-known case of the electric dipole of the moment $\pvec$ is treated employing  the extension as distribution 
of the second order derivative $\d_i\d_j(1/r)$ \cite{Frahm}. In the present paper we make use of the regularized $\delta-$function $\delta_\eps(\rvec),\;\eps\to +\,0$, introduced in Jackson's book \cite{Jackson}:
\beqa\label{2.4}
\delta_\eps(\rvec)=-\frac{1}{4\pi}\Delta\frac{1}{\sqrt{r^2+\eps^2}}=\frac{1}{4\pi}\frac{3\eps^2}{(r^2+\eps^2)^{5/2}}:\nonumber\\ \left\langle\delta_\eps(\rvec),\,\phi(\rvec)\right\rangle=\int\rmd^3x\,\delta_\eps(\rvec)\phi(\rvec)\stackrel{\eps\to 0}{\longrightarrow}\,\phi(0)=\left\langle\,\delta,\,\phi\right\rangle\ .
\eeqa
The test function $\phi(\rvec)$ is an arbitrary element of the space of the Dirac function $\delta$ which can be considered as the subspace of continuous and differentiable functions from $\mathcal{L}^2$.\\
The second order derivative $\d_i\d_j(1/r)$ can be treated by a  procedure of isolating the $\delta-$singularities 
suggested by equation  (2) from Ref.\ \cite{Hnizdo}, writing the regularized  distribution
\beqa\label{2.5}
\fl\;\;\;\;\;\;\d_i\d_j\frac{1}{\sqrt{r^2+\eps^2}}&=&\frac{3x_ix_j}{(r^2+\eps^2)^{5/2}}-\frac{\delta_{ij}}{(r^2+\eps^2)^{3/2}}\nonumber\\
\fl&=&\frac{3x_ix_j-r^2\delta_{ij}}{(r^2+\eps^2)^{5/2}}-\frac{\eps^2\delta_{ij}}{(r^2+\eps^2)^{5/2}}=\frac{3x_ix_j-r^2\delta_{ij}}{(r^2+\eps^2)^{5/2}}-\frac{4\pi}{3}\delta_{ij}\delta_\eps(\rvec)\ .
\eeqa
Denoting $(D)_{(0)}$ a distribution with point-like support, we can write in the present case
\beqa\label{2.6}
\left(\d_i\d_j\frac{1}{r}\right)_{(0)}=-\frac{4\pi}{3}\delta_{ij}\delta(\rvec)\ .
\eeqa
This result, introduced in the expression of the electric point-like dipole field, becomes
\beqa\label{2.7}
\left(\Evec^{(1)}\right)_{(0)}=-\frac{1}{3\eps_0}\,\pvec\,\delta(\rvec)\ .
\eeqa
Let be  the regularized expression of the field $\Evec^{(2)}$ of the point-like quadrupole:
\beqa\label{2.12a}
\left(\Evec^{(2)}(\rvec)\right)_{reg} = -\frac{1}{8\pi\eps_0}\,\evec_i\psf_{jk}\d_i\d_j\d_k\frac{1}{r_\eps} \ ,
\eeqa
where, for simplifying the  notation, it is  introduced 
\beqa\label{2.8}
r_\eps=\sqrt{r^2+\eps^2}\ .
\eeqa
A totally symmetric $n-th$ order tensor can be projected on the subspace of the totally symmetric and trace free ({\bf STF}) tensors. In the particular case $n=2$, this projection is realized writing the decomposition 
\beqa\label{2.13a}
\psf_{ij}=\pc_{ij}+\Lambda\,\delta_{ij}
\eeqa
and choosing the parameter $\Lambda$ such that $\pc_{ii}=0$, i.e.
\beqan
\Lambda=\frac{1}{3}\psf_{ll}\ .
\eeqan
Inserting equation \eref{2.13a} in the regularized expression \eref{2.12a}, we can write
\label{2.13}
\beqan
\left(\Evec^{(2)}(\rvec)\right)_{reg}=-\frac{1}{8\pi\eps_0}\evec_i\,\pc_{jk}\d_i\d_j\d_k\frac{1}{r_\eps}+\frac{1}{2\eps_0}\Lambda\,\evec_i\d_i\delta_{\eps}(\rvec)\ .
\eeqan
One of the singularities of the field is generated by the last term from the previous equation. We have to search the singularities of the contraction of the {\bf STF} tensor $\pc_{ij}$ with the derivative tensor.
The regularized expression of the third-order derivative is given by
\beqan
\fl\d_i\d_j\d_k\frac{1}{r_\eps}=-\frac{15\,x_ix_jx_k}{r^7_\eps}+\frac{3(x_i\delta_{jk}+x_j\delta_{ik}+x_k\delta_{ij})}{r^5_\eps}=-\frac{15\,x_ix_jx_k}{r^7_\eps}+\frac{3\,\delta_{\{ij}\,x_{k\}}}{r^5_\eps}\ .
\eeqan
The notation $\{i_1\,\dots\,i_n\}$ symbolizes the sum over all the transpositions of the indexes $i_1\dots i_n$ 
which correspond to distinct terms. Applying the same procedure as in the case of equation \eref{2.5},
\beqa\label{2.9}
\fl\;\;\;\;\;\d_i\d_j\d_k\frac{1}{r_\eps}=-\frac{15\,x_ix_jx_k}{r^7_\eps}
+\frac{3\,\delta_{\{ij}x_{k\}}}{r^5_\eps}
=\frac{-15\,x_ix_jx_k+3r^2\delta_{\{ij}x_{k\}}}{r^7_\eps}+\frac{3\eps^2\,\delta_{\{ij}x_{k\}}}{r^7_\eps}\ .
\eeqa
Writing the partial derivative of  $\delta_\eps(\rvec)$, we obtain
\label{2.10}
\beqan
\frac{\eps^2\,x_i}{r^7_\eps}=-\frac{4\pi}{15}\,\d_i\delta_\eps(\rvec) \ ,
\eeqan
which, inserted in equation \eref{2.9}, gives
\beqa\label{2.11}
\d_i\d_j\d_k\frac{1}{r_\eps}=\frac{-15\,x_ix_jx_k+3r^2\delta_{\{ij}x_{k\}}}{r^7_\eps}-\frac{4\pi}{5}\delta_{\{ij}\d_{k\}}\delta_\eps(\rvec) \ ,
\eeqa
such that the $\delta-$singularities of the third-order derivative are given by
\beqa\label{2.12}
\left(\d_i\d_j\d_k\frac{1}{r}\right)_{(0)}=-\frac{4\pi}{5}\delta_{\{ij}\d_{k\}}\delta(\rvec)\ ,
\eeqa
a well-known result \cite{Frahm}. Therefore, the singularity of the contraction of the {\bf STF} moment with the derivative tensor is given by
\beqan
\left(\pc_{jk}\d_i\d_j\d_k\frac{1}{r}\right)_{(0)}=-\frac{8\pi}{5}\pc_{ij}\d_j\delta(\rvec)\ .
\eeqan
Finally, for the point-like quadrupole, we obtain for the electric field
\beqan
\left(\Evec^{(2)}(\rvec)\right)_{(0)}=\evec_i\left[\frac{1}{5\eps_0}\,\pc_{ij}\d_j\delta(\rvec)+\frac{1}{2\eps_0}\Lambda\,\d_i\delta(\rvec)\right] \ ,
\eeqan
or, using a tensorial notation,
\beqa\label{2.16}
\left(\Evec^{(2)}(\rvec)\right)_{(0)}=\frac{1}{5\eps_0}\pct^{(2)}\vert\vert\nablav\delta(\rvec)+\frac{1}{2\eps_0}\Lambda\nablav\delta(\rvec)\ .
\eeqa
In the last equation, it is employed the general notation $\Tsft^{(n)}$ for the $n-th$ order tensor and, also, the notation for the tensor contraction:
\beqan
 ({\Atens}^{(n)}||{\Btens}^{(m)})_{i_1 \cdots i_{|n-m|}}
=\left\{\begin{array}{ll}
A_{i_1 \cdots i_{n-m}j_1 \cdots j_m}B_{j_1 \cdots j_m} & ,\; n>m\\
A_{j_1 \cdots j_n}B_{j_1 \cdots j_n} & ,\; n=m\\
A_{j_1 \cdots j_n}B_{j_1 \cdots j_n i_1 \cdots i_{m-n}} & ,\; n<m
\end{array} \right.\ .
\eeqan
Searching directly the singularities associated to distributions defined as contractions of electric (or magnetic) moments and derivative tensors represents the basic procedure adopted in the present paper. This procedure implies an appreciable simplicity of the calculation especially for the higher-order multipoles.\\
We point out the invariance of the electrostatic field  for $\left(\Evec^{(2)}(\rvec)\right)_{r\ne0}$  to the substitution of the ``primitive `` moment $\psft^{(2)}$ by the {\bf STF} one, $\psft^{(2)}\to\pct^{(2)}$, such that
\beqa\label{2.19a}
 \left(\Evec^{(2)}(\rvec)\right)_{r\ne0}=-\frac{1}{8\pi\eps_0}\pct^{(2)}\vert\vert\nablav^3\frac{1}{r}\ .
\eeqa
Obviously, searching the $\delta-$singularities of $\Evec^{(2)}$ starting from this last expression, the singular term containing the parameter $\Lambda$ from equation \eref{2.16} is lost. The primitive tensor $\psft^{(2)}$ includes all the necessary elements for finding the $\delta-$singularities of the field. Employing, for example, the multipole expansion in terms of the spherical functions $Y_{lm}(\theta,\varphi)$, it is equivalent to employing the expansion \eref{2.19a} yielding therefore,  a wrong result for such type of singularities.\\
Let us consider the regularized expression of the electric field $\Evec^{(3)}(\rvec)$:
\beqa\label{2.20}
\left(\Evec^{(3)}\right)_{reg}=\frac{1}{24\pi\eps_0}\psft^{(3)}\vert\vert\nablav^4\frac{1}{r_\eps}
=\frac{1}{24\pi\eps_0}\evec_i\,\psf_{jkl}\,\d_i\d_j\d_k\d_l\frac{1}{r_\eps}\ .
\eeqa
The {\bf STF} projection of the tensor $\psft^{(3)}$ is introduced by the following decomposition:
\beqa\label{2.21}
\psf_{ijk}=\pc_{ijk}+\Lambda_{\{i}\delta_{jk\}}\ .
\eeqa
The parameters $\Lambda_i$ are established requiring the vanishing of all the traces of the symmetric tensor $\psft^{(3)}$:
\beqan
\psf_{ijj}=0,\;(i=1,\,2,\,3)\ .
\eeqan
The results are given by
\beqa\label{2.22}
\Lambda_i=\frac{1}{5}\,\psf_{ijj}\ .
\eeqa
The insertion of equation \eref{2.21} in equation  \eref{2.20} gives
\beqa\label{2.23}
\left(\Evec^{(3)}\right)_{reg}&=&\frac{1}{24\pi\eps_0}\left[\pct^{(3)}\vert\vert\nablav^4\frac{1}{r_\eps}
+\frac{3}{24\pi\eps_0}\evec_i\Lambda_j\d_i\d_j\Delta\frac{1}{r_\eps}
\right]\nonumber\\
&=&\frac{1}{24\pi\eps_0}\pct^{(3)}\vert\vert\nablav^4\frac{1}{r_\eps}-\frac{1}{2\eps_0}\evec_i\Lambda_j\d_j\d_i\delta_\eps(\rvec)\ .
\eeqa
The fourth-order derivative of $1/r$ is given by 
\beqan
\d_i\d_j\d_k\d_l\frac{1}{r}=\frac{7!!\,x_ix_jx_kx_l}{r^9}-\frac{5!!\,\delta_{\{ij}\,x_kx_{l\}}}{r^7}+\frac{3\,\delta_{\{ij}\,\delta_{kl\}}}{r^5}\ .
\eeqan
We are interested only in writing the contraction $\pc_{jkl}\d_i\d_j\d_k\d_l(1/r)$ and:
\beqan
\pc_{jkl}\d_i\d_j\d_k\d_l\frac{1}{r}=\pc_{jkl}\left(\frac{7!!\,x_ix_jx_kx_l}{r^9}-\frac{5!!\,\delta_{\{ij}\,x_kx_{l\}}}{r^7}\right) \ ,
\eeqan
since $\pc_{jkl}\delta_{\{ij}\,\delta_{kl\}}=0$.
It is easy to see that the same formula applies to $1/r_\eps$ by the simple substitution $r\to r_\eps$.
Writing the corresponding regularized expression and applying the same procedure as in the case of the third-order derivative for obtaining equation \eref{2.11}, 
we write
\beqa\label{2.24}
\fl\;\;\;\;\;\;\;\pc_{jkl}\d_i\d_j\d_k\d_l\frac{1}{r_\eps}=\pc_{jkl}\left(\frac{7!!\,x_ix_jx_kx_l-5!!\,r^2\,\delta_{\{ij}\,x_kx_{l\}}}{r^9_\eps}
- 15\frac{\eps^2\,\delta_{\{ij}\,x_kx_{l\}}}{r^9_\eps}\right)\ .
\eeqa
Expressing the second partial derivative of $\delta_\eps(\rvec)$, one obtains
\beqan
\frac{\eps^2\,x_ix_j}{r^9_\eps}=\frac{4\pi}{7!!}\d_i\d_j\,\delta_\eps(\rvec)+\frac{\eps^2\,\delta_{ij}}{7\,r^7_\eps}\ .
\eeqan
The insertion of this equation in the last fraction from equation \eref{2.24} gives
\beqan
\fl\;\;\;\;\;\;\;\;\pc_{jkl}\d_i\d_j\d_k\d_l\frac{1}{r_\eps}=\pc_{jkl}\left(\frac{7!!\,x_ix_jx_kx_l-5!!\,r^2\,\delta_{\{ij}\,x_kx_{l\}}}{r^9_\eps}
-\frac{4\pi}{7}\delta_{\{ij}\,\d_k\d_{l\}}\delta_\eps(\rvec)\right)\ .
\eeqan
From the last result, we identify the $\delta-$singularity:
\beqa\label{2.26}
\fl\;\;\;\;\;\;\;\;\left(\pc_{jkl}\,\d_i\d_j\d_k\d_l\frac{1}{r}\right)_{(0)}&=&-\frac{4\pi}{7}\pc_{jkl}\left(\delta_{ij}\d_k\d_l\,\delta(\rvec)+
\delta_{ik}\d_j\d_l\,\delta(\rvec)+\delta_{il}\d_j\d_k\,\delta(\rvec)\right)\nonumber\\
\fl\;\;\;\;\;\;\;\;&=&-4\pi\,\frac{3}{7}\pc_{ijk}\d_j\d_k\,\delta(\rvec) \ ,
\eeqa
where the vanishing of the contraction of $\pc_{jkl}$ with $\delta_{jk},\,\delta_{jl},\,\delta_{kl}$ and the symmetry properties of the $\pct^{(3)}$ components are considered. With tensorial notation,
\beqa\label{2.27}
\left(\pct^{(3)}\vert\vert\nablav^4\frac{1}{r}\right)_{(0)}=-4\pi\,\frac{3}{7}\pct^{(3)}\vert\vert\nablav^2\delta(\rvec)\ .
\eeqa
This last result together with equation \eref{2.23} and the limit for $\eps\to0$ in the distribution space leads to the $\delta-$singularity of $\Evec^{(3)}$:
\beqa\label{2.28}
\left(\Evec^{(3)}(\rvec)\right)_{(0)}=-\frac{1}{14\,\eps_0}\,\pct^{(3)}\vert\vert\nablav^2\delta(\rvec)
-\frac{1}{2\,\eps_0}\latens\vert\vert\nablav^2\delta(\rvec) \ ,
\eeqa
where $\latens=\Lambda_i\evec_i$.\\
This procedure can be generalized to any arbitrary higher-order $n$.

\section{  Magnetostatic field}
We consider  the first three terms from the magnetostatic   field expansions \cite{Jackson}, \cite{Castell}:
\beqa\label{3.2}
\fl\Bvec(\rvec)=&\frac{\mu_0}{4\pi}\,\left(\nablav^2\vert\vert\frac{\mvec}{r}-\Delta\frac{\mvec}{r}-\frac{1}{2}\nablav^3\vert\vert\frac{\msft^{(2)}}{r}+\frac{1}{2}\nablav\vert\vert\Delta\frac{\msft^{(2)}}{r}
+\frac{1}{6}\nablav^4\vert\vert\frac{\msft^{(3)}}{r}\right.\nonumber\\ 
\fl&\left.
-\frac{1}{6}\nablav^2\vert\vert\Delta\frac{\msft^{(3)}}{r}
\,+\,\dots\right)
=\frac{\mu_0}{4\pi}\evec_i\left(\d_i\d_j\frac{m_j}{r}-\Delta\frac{m_i}{r}-\frac{1}{2}\d_i\d_j\d_k\frac{\msf_{jk}}{r}\right.\nonumber\\
\fl&+\left.\frac{1}{2}\d_j\Delta\frac{\msf_{ji}}{r}+\frac{1}{6}\d_i\d_j\d_k\d_l\frac{\msf_{jkl}}{r}-\frac{1}{6}\d_j\d_k\Delta\frac{\msf_{jki}}{r}\,+\dots\right)\ .
\eeqa
The magnetic moments are defined by \cite{Jackson}, \cite{Castell}
\beqa\label{3.3}
 m_i=\frac{1}{2}\int_{\dom}\rmd^3x\,\left(\rvec\times\jvec\right)_i,\;\;
\msf_{ik}=\frac{2}{3}\int_{\dom}\rmd^3x\,x_i\,\left(\rvec\times\jvec\right)_k,\nonumber\\
\msf_{ijk}=\frac{3}{4}\int_{\dom}
\rmd^3x\,x_ix_j\left(\rvec\times\jvec\right)_k\ .
\eeqa
Let us write the regularized expression of the dipolar magnetic field:
\beqan
\left(\Bvec(\rvec)\right)_{reg}=\frac{\mu_0}{4\pi}\left(\evec_i\d_i\d_j\frac{m_j}{r}+4\pi\mvec\,\delta_\eps(\rvec)\right)\ .
\eeqan
Inserting equations \eref{2.5} and \eref{2.6}, we obtain the following result for the singular term
\beqa\label{3.4}
\left(\Bvec(\rvec)\right)_{(0)}=-\frac{\mu_0}{3}\mvec\,\delta(\rvec)+\mu_0\,\mvec\,\delta(\rvec)=\frac{2\mu_0}{3}\,\mvec\,\delta(\rvec)\ .
\eeqa
Comparing equations \eref{2.7} and \eref{3.4}, it is seen the difference between the two expressions concerning the proportional factor associated to the dipolar moment $\pvec$ or $\mvec$. Formally, this difference is due to the supplementary term $\Delta(\mvec/r)$  in the magnetic case.  Physically, the difference becomes easily obvious if one considers the fictitious magnetic shells (or sheets) employed in the Amp\`{e}re formalism. It suffices to consider the case of the point-like magnetic dipole which can be taken as the limit of a current loop of infinitesimal size. For finite dimensions, the field of the loop can be derived from a scalar potential $\Phi_m$ which is defined by an integral on the corresponding sheet and having a ``jump'' in all their points. Just this jump generates the $\delta$-singularity corresponding to the second term from equation \eref{3.4}. An explicit calculation of this limit when the loop concentrates in a point is given in Ref.\ 
\cite{Corbo}.\\
Concerning the higher-order magnetic multipoles, it is necessary to specify the procedure of projecting  the corresponding moments on the {\bf STF} tensor subspace. In the case of the quadrupolar magnetic moment, the components $\msf_{ij}$ are defined in equation \eref{3.3}. These components, not being symmetric in the two indexes, will involve two steps in establishing the {\bf STF} tensor $\mct^{(2)}=\Tcal(\msft^{(2)})$, where by $\Tcal$ is denoted the correspondence between an arbitrary tensor and its {\bf STF} projection. We establish firstly the symmetric projection denoted here by $\mlrt^{(2)}$ and, secondly, we apply the  known procedure of establishing the {\bf STF} projection of this one. Let us  write the identity 
\beqa\label{3.5}
\msf_{ij}=\frac{1}{2}\left(\msf_{ij}+ \msf_{ji}\right)+ \frac{1}{2}\left(\msf_{ij}- \msf_{ji}\right)=
\mlr_{ij}+\frac{1}{2}\left(\msf_{ij}- \msf_{ji}\right)\ .
\eeqa
 In this case ($n=2$),  $\mct^{(2)}=\mlrt^{(2)}$ since $\msf_{jj}=0$  and, consequently, corresponds to the {\bf STF} projection. Therefore,
  \beqa\label{3.6}    
\msf_{ij}=\mc_{ij}+\frac{1}{2}\left(\msf_{ij}- \msf_{ji}\right)\ .
 \eeqa
 Let be the regularized expression of the 4-polar magnetic field:
\beqan    
\left(\Bvec^{(2)}(\rvec)\right)_{reg}=\frac{\mu_0}{8\pi}\evec_i\,\left[-\msf_{jk}\d_i\d_j\d_k\frac{1}{r_\eps}
+\msf_{ji}\d_j\Delta\frac{1}{r_\eps}\right]
\eeqan
and inserting equations \eref{2.4} and \eref{3.6},
\beqan
\left(\Bvec^{(2)}(\rvec)\right)_{reg}=-\,\frac{\mu_0}{8\pi}\evec_i\,\left[\mc_{jk}\,\d_i\d_j\d_k\frac{1}{r_\eps}
+4\pi\msf_{ji}\d_j\delta_\eps(\rvec) 
\right]\ ,
\eeqan 
where the vanishing of the contraction $\left(\msf_{jk}-\msf_{kj}\right)\d_i\d_j\d_k$ is considered.
 It remains to separate the part generating the $\delta-$singularity of the expression $\mc_{jk}\,\d_i\d_j\d_k(1/r_\eps)$ from the last equation. This objective is performed in the same manner as in the case of the electric field $\Evec^{(2)}$, employing equation \eref{2.12}. We obtain the final result
  \beqa\label{3.7} 
\left(\Bvec^{(2)}(\rvec)\right)_{(0)}=\mu_0\left(\frac{1}{5}\mct^{(2)}\vert\vert\nablav\delta(\rvec)-\frac{1}{2}\nablav\delta(\rvec)\vert\vert\msft^{(2)}\right)\ .   
\eeqa
This result can be written also in terms of irreducible tensors, but this is only an optional problem, the main objective  of expressing the $\delta-$singularities of the field being realized by the result \eref{3.7}. If we introduce the vector $\bbox{N}$ defined by the components
 \beqa\label{3.8} 
N_i=\eps_{ijk}\msf_{jk}=\frac{2}{3}\int_\dom\,\rmd^3x\,\left[\rvec\times\left(\rvec\times\jvec\right)\right]_i \ ,
\eeqa
equation \eref{3.6} can be written as
 \beqa\label{3.9} 
 \msf_{ij}=\mc_{ij}+\frac{1}{2}\eps_{ijk}N_k \ .
 \eeqa
The insertion of equation \eref{3.9} in equation \eref{3.7} gives
 \beqa\label{3.10} 
\left(\Bvec^{(2)}(\rvec)\right)_{(0)}=-\,\frac{3\mu_0}{10}\,\mct^{(2)}\vert\vert\nablav\delta_\eps(\rvec)-\frac{\mu_0}{4}\bbox{N}\times\nablav\delta_\eps(\rvec)\ .
\eeqa
Beginning from $n=3$, the symmetric projection of the magnetic moment is not the same with the {\bf STF} projection and, consequently, the second step of the procedure becomes an actual calculation. Let us write the regularized expression of the $3-rd$ order magnetic field:
 \beqa\label{3.11} 
\left(\Bvec^{(3)}(\rvec)\right)_{reg}=\frac{\mu_0}{24\pi}\nablav^4\vert\vert\frac{\msft^{(3)}}{r_\eps}+\frac{\mu_0}{6}\nablav^2\delta_\eps(\rvec)\vert\vert\msft^{(3)}\ .
\eeqa
Writing the identity
 \beqa\label{3.12} 
 \msf_{ijk}&=\frac{1}{3}\left(\msf_{ijk}+\msf_{kji}+\msf_{ikj}\right)
 +\frac{1}{3}\left[\left(\msf_{ijk}-\msf_{kji}\right)+\left(\msf_{ijk}-\msf_{ikj}\right)\right]\nonumber\\
 &=\mlr_{ijk}+\frac{1}{3}\left[\left(\msf_{ijk}-\msf_{kji}\right)+\left(\msf_{ijk}-\msf_{ikj}\right)\right] \ ,
\eeqa
we introduce the {\bf STF} projection $\mct^{(3)}$ of the symmetric tensor $\mlrt^{(3)}$ by the equation
 \beqa\label{3.13} 
 \mc_{ijk}=\mlr_{ijk}+\delta_{\{ij}\widetilde{\Lambda}_{k\}},\;\;\widetilde{\Lambda}_i=\frac{1}{5}\mlr_{qqi}
 =\frac{1}{15}\msf_{qqi}\ .
\eeqa
Employing equation \eref{3.12} and since $\left[\left(\msf_{ijk}-\msf_{kji}\right)+\left(\msf_{ijk}-\msf_{ikj}\right)\right]\d_i\d_j\d_k\d_l(1/r_\eps)=0$, we can write
\beqan
\nablav^4\vert\vert\frac{\msft^{(3)}}{r_\eps}=\nablav^4\vert\vert\frac{\mlrt^{(3)}}{r_\eps} \ .
\eeqan
 Instead of equation \eref{3.11}, we will have
\beqan
\left(\Bvec^{(3)}(\rvec)\right)_{reg}=\frac{\mu_0}{24\pi}\nablav^4\vert\vert\frac{\mlrt^{(3)}}{r_\eps}
+\frac{\mu_0}{6}\nablav^2\delta_\eps(\rvec)\vert\vert\msft^{(3)}\ .
\eeqan
The first term from the right-hand side of the above equation can be processed as in the case of $\Evec^{(3)}$, equation \eref{2.20}, employing the result given by equation \eref{2.28} with $\eps_0\to 1/\mu_0$ and $\Lambda\to \widetilde{\Lambda}$ such that, finally,
 \beqa\label{3.14} 
\fl\;\;\;\;\;\;\;\;\left(\Bvec^{(3)}(\rvec)\right)_{(0)}=-\frac{\mu_0}{14}\mct^{(3)}\vert\vert\nablav^2\delta(\rvec)-\frac{\mu_0}{2}\widetilde{\Lambda}\vert\vert\nablav^2\delta(\rvec)+\frac{\mu_0}{6}\nablav^2\delta(\rvec)\vert\vert\msft^{(3)}\ .
\eeqa

\section{General formulas - a mathematical digression}
The general expansions of the static electric and magnetic fields are given by
 \beqa\label{4.01}   
\Evec(\rvec)=\frac{1}{4\pi\eps_0}\suml_{n\ge0}\frac{(-1)^{n-1}}{n!}\psft^{(n)}\vert\vert\nablav^{n+1}\frac{1}{r}\ 
\eeqa
and
 \beqa\label{4.02} 
 \Bvec(\rvec)=\frac{\mu_0}{4\pi}\suml_{n\ge1}\frac{(-1)^{n-1}}{n!}\left[\nablav^{n+1}\vert\vert\frac{\msft^{(n)}}{r}
 -\nablav^{n-1}\vert\vert\Delta\frac{\msft^{(n)}}{r}\right]   \ .
\eeqa
The general definition of the electric $n-th$ order moment is given by
 \beqan 
\psft^{(n)}=\int_\dom\rmd^3x\,\,\rvec^n\,\rho(\rvec):\;\;\psf_{i_1\dots i_n}=\int_\dom\,\rmd^3x\,x_{i_1}\dots x_{i_n}\,\rho(\rvec)\ ,
\eeqan
and for the magnetic $n-th$ order moment by \cite{Castell}
 \beqan 
\msft^{(n)}=\frac{n}{n+1}\int_\dom\rmd^3x\,\rvec^n\times\jvec:\;\;\msf_{i_1\dots i_n}=\int_\dom\rmd^3x\,x_{i_1}\dots x_{i_{n-1}}\left(\rvec\times\jvec\right)\ .
\eeqan
 In Ref.\ \cite{Estrada}, a formula for expressing the ordinary $n-th$ order derivative of $1/r^m$ is given (equation (4.16) from this reference). With our notation, 
  \beqa\label{4.1}   
\d_{i_1}\dots\d_{i_n}\frac{1}{r^m}&=& \suml^{[\frac{n}{2}]}_{k=0}
\frac{(-1)^{n-k}m(m+2)\dots(2n-2k+m-2)}{r^{2n-2k+m}}\nonumber\\
&&\times \,\delta_{\{i_1i_2}\dots\delta_{i_{2k-1}\,i_{2k}}x_{2k+1}\dots x_{i_n\}} \ ,
\eeqa
where $[\al]$ is the integer part of $\al$. This formula can be employed for calculating the derivatives of 
$1/r_\eps$ by the simple substitution $r\to r_\eps$ in equation \eref{4.1}. We are interested of the case m=1:
 \beqa\label{4.2}   
\fl\;\;\;\;\;\;\;\;\;\;\d_{i_1}\dots\d_{i_n}\frac{1}{r_\eps}=\suml^{[\frac{n}{2}]}_{k=0}\frac{(-1)^{n-k}(2n-2k-1)!!}{r^{2n-2k+1}_\eps}\delta_{\{i_1i_2}\dots\delta_{i_{2k-1}i_{2k}}x_{i_{2k+1}}\dots x_{i_n\}}\ .
\eeqa
We  also need the formula for the derivatives of $\delta_\eps(\rvec)$:
 \beqa\label{4.3}   
\fl\;\;\;\;\;\;\d_{i_1}\dots\d_{i_n}\delta_\eps(\rvec)=\frac{\eps^2}{4\pi}\suml^{[\frac{n}{2}]}_{k=0}\frac{(-1)^{n-k}(2n-2k+3)!!}
{r^{2n-2k+5}_\eps}\delta_{\{i_1i_2}\dots\delta_{i_{2k-1}i_{2k}}x_{i_{2k+1}}\dots x_{i_n\}}\ .
\eeqa
Equations  \eref{4.2} and \eref{4.3} will be employed in the following for expressing the singular terms with point-like support of the electric and magnetic fields for arbitrary $n$.\\
Let us firstly take the regularized expression of the $n-th$ order multipole electric field 
 \beqa\label{4.4} 
\fl\;\;\;\;\;\;\left(\Evec^{(n)}(\rvec)\right)_{reg}=\frac{(-1)^{n-1}\,}{4\pi\eps_0\,n!}\psft^{(n)}\vert\vert\nablav^{n+1}\frac{1}{r_\eps}
 =\frac{(-1)^{n-1}\,}{4\pi\eps_0\,n!}\,\evec_i\,\psf_{i_1\dots i_n}\,\d_i\,\d_{i_1}\dots \d_{i_n}\frac{1}{r_\eps}\ .
\eeqa
The {\bf STF} projection of the tensor $\psft^{(n)}$ is realized by an obvious generalization of equation \eref{2.21}:
 \beqa\label{4.5}
 \psf_{i_1\dots i_n}=\pc_{ i_1\dots i_n}+\delta_{\{i_1i_2}\,\Lambda_{i_3\dots i_n\}} \ ,
 \eeqa
where $\Lambda^{(n-2)}$ is a totally symmetric tensor. This last tensor has a general expression given in Refs.  \cite{Thorne} and \cite{App} (the detracer theorem) which, with our notation, is written as
 \beqan
 \fl\;\;\;\;\;\;\lasf_{i_1\dots i_{n-2}}\big[\psft^{(n)}\big]=
\suml^{[n/2-1]}_{m=0}\frac{(-1)^m[2n-1-2(m+1)]!!}{(m+1)(2n-1)!!}
\delta_{\{i_1i_2}\dots \delta_{i_{2m-1}i_{2m}}\psf^{(n:\,m+1)}_{i_{2m+1}\dots i_{n-2}\}}\ .
\eeqan
$\psf^{(n:m)}_{i_{2m+1}\dots i_n}$ denotes the components of the $(n-2\,m)-th$ order tensor 
obtained from $\psft^{(n)}$ by contracting $m$ pairs of symbols $i$. This theorem, though not employed in the present paper, is given for the  reader interested in  extending these calculation for higher orders.\\ 
Further, we insert formula \eref{4.5} in equation \eref{4.4} and we focus on the calculation of the  contraction of the electric moment tensor with the derivative one:
\beqan
\fl &&\psft^{(n)}\vert\vert\nablav^{n+1}\frac{1}{r_\eps}=\evec_i\,\psf_{i_1\dots i_n}\,\d_i\,\d_{i_1}\dots \d_{i_n}\frac{1}{r_\eps}=\pct^{(n)}\vert\vert\nablav^{n+1}\frac{1}{r_\eps}\\
\fl\;\;\;\;\;\;&+&\evec_i\,\d_i\,\delta_{\{i_1i_2}\,\Lambda_{i_3}\dots i_{n\}}\d_{i_1}\dots\d_{i_n}\frac{1}{r_\eps}
=\pct^{(n)}\vert\vert\nablav^{n+1}\frac{1}{r_\eps}+\frac{n(n-1)}{2}\evec_i\d_i\,\d_{i_1}\dots\d_{i_{n-2}}\Delta\frac{1}{r_\eps} \ ,
\eeqan
where the symmetry of the tensors $\latens^{(n-2)}$ and $\nablav^n$ is considered. It is realized a first separation of  $\delta-$singularities set by the introduction of the $\delta_\eps$ function:
 \beqa\label{4.7}
\psft^{(n)}\vert\vert\nablav^{n+1}\frac{1}{r_\eps}&=&\pct^{(n)}\vert\vert\nablav^{n+1}\frac{1}{r_\eps}
-4\pi\frac{n(n-1)}{2}\evec_i\,\d_i\,\d_{i_1}\dots\d_{i_{n-2}}\,\delta_\eps(\rvec)\nonumber\\
&=&\pct^{(n)}\vert\vert\nablav^{n+1}\frac{1}{r_\eps}-4\pi\frac{n(n-1)}{2}\,\latens^{(n-2)}\vert\vert\nablav^{n-1}\delta_\eps(\rvec)\ .
\eeqa
Employing  the formula \eref{4.2}, we can write
\beqan
\fl\pct^{(n)}\vert\vert\nablav^{n+1}\frac{1}{r_\eps}=-\evec_{i_{n+1}}\,\pc_{i_1\dots i_n}\d_{i_1}\dots\d_{i_{n+1}}\frac{1}{r_\eps}\\
\fl \;\;\;=-\evec_{i_{n+1}}\,\pc_{i_1\dots i_n}\suml^{[(n+1)/2]}_{k=0}\frac{(-1)^{n-k}(2n-2k+1)!!}{r^{2n-2k+3}_\eps}
\delta_{\{i_1i_2}\dots\delta_{i_{2k-1}i_{2k}}\,x_{i_{2k+1}}\dots x_{i_{n+1\}}}    \\
\fl\;\;\;=-\evec_{i_{n+1}}\,\pc_{i_1\dots i_n}\left[\frac{(-1)^n(2n+1)!!\,x_{i_1}\dots x_{i_{n+1}}}{r^{2n+3}_\eps}\right.
-\left.\frac{(-1)^n(2n-1)!!\,\,\delta_{\{i_1i_2}\,x_{i_3}\dots x_{i_{n+1}\}}}{r^{2n+1}_\eps}\right]\ ,
\eeqan
since the contractions 
\beqan
\pc_{i_1\dots i_n}\,\delta_{\{i_1i_2}\dots\delta_{i_{2k-1}}\delta_{i_{2k}}\,x_{i_3}\dots x_{i_{n+1}}\}
\eeqan
vanish for $k\ge 2$. We perform the separation  of 
the singular part writing:
 \beqa\label{4.8} 
 \fl\pct^{(n)}\vert\vert\nablav^{n+1}\frac{1}{r_\eps}&=&(-1)^{n-1}\evec_{i_{n+1}}\pc_{i_1\dots i_n}\nonumber\\
 &&\times\left[\frac{(2n+1)!!\,x_{i_1}\dots i_{n+1}-(2n-1)!!\,r^2\,\delta_{\{i_1i_2}\,x_{i_3}\dots x_{i_{n+1}\}}}
 {r^{2n+3}_\eps}\right.\nonumber\\
 &&-\left.\frac{(2n-1)!!\,\eps^2\delta_{\{i_1i_2}x_{i_3}\dots x_{i_{n+1}\}}}{r^{2n+3}_\eps}
 \right]\ .
 \eeqa
Equation  \eref{4.3} gives the following relation:
 \beqa\label{4.9} 
\fl -\frac{\eps^2\,x_{i_1}\dots x_{i_{n-1}}}{r^{2n+3}_\eps}&=&-\frac{4\pi\,(-1)^n}{(2n+1)!!}\,\d_{i_1}\dots\d_{i_{n-1}}\,\delta_\eps(\rvec)\\
 \fl&-&\frac{(-1)^n}{(2n+1)!!}\suml^{[(n-1)/2]}_{k=1}\frac{(2n-2k+1)!!}{r^{2n-2k+3}_\eps}
 \,\delta_{\{i_1i_2}\dots\delta_{i_{2k-1}i_{2k}}x_{i_{2k+1}}\dots x_{i_{n-1}\}}\nonumber \ .
 \eeqa
 Inserting equation \eref{4.9} in equation \eref{4.8}, we obtain
  \beqa\label{4.10}
 \fl\;\;\;\;&&\pct^{(n)}\vert\vert\nablav^{n+1}\frac{1}{r_\eps}\nonumber\\
 \fl\;\;\;\;&=&(-1)^{n-1}\evec_{i_{n+1}}\pc_{i_1\dots i_n}
 \left[\frac{(2n+1)!!\,x_{i_1}\dots i_{n+1}-(2n-1)!!\,r^2\,\delta_{\{i_1i_2}\,x_{i_3}\dots x_{i_{n+1}\}}}
 {r^{2n+3}_\eps}\right]\nonumber\\
\fl\;\;\;\; &&+\suml_{k\ge 1}(\dots)-\frac{4\pi}{2n+1}\,\evec_{i_{n+1}}\pc_{i_1\dots i_n}\delta_{\{i_1i_2}\,\d_{i_3}\dots\d_{i_{n+1}\}}\,\delta_\eps(\rvec)\ .
\eeqa
In the above equation, the symbol $\sum_{k\ge 1}(\dots)$ represents the contraction of $\pc_{i_1\dots i_n}$ with a sum in which each term contains a Kronecker symbols product with at least two factors, one of these factors being  $\delta_{i_j,i_k}$ with $j,\,k\,\le n$. Consequently, all these contractions are vanishing. With this result, we can express the $\delta-$singularity from equation \eref{4.10}:
 \beqa\label{4.11}
 \left(\pct^{(n)}\vert\vert\nablav^{n+1}\frac{1}{r}\right)_{(0)}&=&-\frac{4\pi}{2n+1}\evec_i\,\pc_{i_1\dots i_n}\, \delta_{\{ii_1}\,\d_{i_2}\dots\d_{i_n\}}\delta(\rvec)\nonumber\\
 &=&-\,4\pi\frac{\,n}{2n+1}\,\pct^{(n)}\vert\vert\nablav^{n-1}\delta(\rvec)
 \ .
 \eeqa
 From equations \eref{4.4}, \eref{4.7} and \eref{4.11}, we can write the $\delta-$singularity of $\Evec^{(n)}$:
  \beqa\label{4.12}
\left(\Evec^{(n)}(\rvec)\right)_{(0)}=\frac{(-1)^n}{\eps_0\,(n-1)!}\left(\frac{1}{2n+1}\pct^{(n)}+\frac{n-1}{2}\latens^{(n-2)}\right)\vert\vert\nablav^{n-1}\,\delta(\rvec)\ .
\eeqa
Let us consider finally the singularity of the magnetic field.
 The tensor $\msft^{(n)}$  is symmetric only in the first $n-1$ indices    and satisfies the property
 \beqan   
\msf_{i_1\dots i_{n-2}\,qq}=0\ .
\eeqan
In the first step, we must obtain the symmetric projection  $\mlrt^{(n)}$ of the tensor
$\msft^{(n)}$. For $n\ge 3$, we can generalize equation \eref{3.5}  writing the identity:
 \beqa\label{a.16}  
\fl\;\;\;\;\;\msf_{i_1\dots i_n}&=&\frac{1}{n}\left(\msf_{i_1\dots i_n}+\msf_{i_ni_2\dots i_{n-1} i_1}+\dots+ \msf_{i_1\dots i_n i_{n-1}}
\right)\nonumber\\
\fl&&+\frac{1}{n}\left[\left(\msf_{i_1\dots i_n}-\msf_{i_n\dots i_{n-1}i_1}\right)+\dots\left(\msf_{i_1\dots i_n}
-\msf_{i_1\dots i_{n-2}i_n\,i_{n-1}}\right)\right]\nonumber\\
\fl&=&\mlr_{i_1\dots i_n}+\frac{1}{n}\left[\left(\msf_{i_1\dots i_n}-\msf_{i_n\dots i_{n-1}i_1}\right)+\dots\left(\msf_{i_1\dots i_n}
-\msf_{i_1\dots i_{n-2}i_n\,i_{n-1}}\right)\right] \ ,
 \eeqa
 where   $\mlrt^{(n)}$  represents  the symmetric part of the tensor $\msft^{(n)}$.
Let us write the regularized expression of the $n-th$ order multipole magnetic field:
 \beqa\label{4.13}
\fl\;\;\;\;\;\;\left(\Bvec^{(n)}(\rvec)\right)_{reg}&=&\frac{(-1)^{n-1}\mu_0}{4\pi\,n!}\left[\nablav^{n+1}\vert\vert\frac{\msft^{(n)}}{r_\eps}
 -\nablav^{n-1}\vert\vert\Delta\frac{\msft^{(n)}}{r_\eps}\right] \ .
 \eeqa
 The insertion of equation \eref{a.16} in the above equation gives
 \beqa\label{4.14}
\nablav^{n+1}\vert\vert\frac{\msft^{(n)}}{r_\eps}
=\nablav^{n+1}\vert\vert\frac{\mlrt^{(n)}}{r_\eps} \ ,
\eeqa
since 
\beqan
\d_{i_1}\dots\d_{i_n}\frac{1}{r_\eps}\left[\left(\msf_{i_1\dots i_n}-\msf_{i_n\dots i_{n-1}i_1}\right)+\dots\left(\msf_{i_1\dots i_n}
-\msf_{i_1\dots i_{n-2}i_n\,i_{n-1}}\right)\right]=0\ .
\eeqan
Let us denote 
 \beqa\label{4.15}
 \widetilde{\Lambda}^{(n-2)}=\latens\left(\mlrt^{(n)}\right):\;\;\mlr_{i_1\dots i_n}=
 \mc_{i_1\dots i_n}+\delta_{\{i_1i_2}\widetilde{\Lambda}_{i_3\dots i_n\}}\ .
 \eeqa
With this notation,
\beqan
\nablav^{n+1}\vert\vert\frac{\msft^{(n)}}{r_\eps}&=&\nablav^{n+1}\vert\vert\frac{\mct^{(n)}}{r_\eps}
+\evec_i\,\d_i\d_{i_1}\dots\d_{i_n}\,\delta_{\{i_1i_2}\,\widetilde{\Lambda}_{i_3\dots i_n\}}\,\frac{1}{r_\eps}\\
&=&\nablav^{n+1}\vert\vert\frac{\mct^{(n)}}{r_\eps}
+\frac{n(n-1)}{2}\evec_i\widetilde{\Lambda}_{i_1\dots i_{n-2}}\d_i\,\d_{i_1}\dots\d_{i_{n-2}}\,\Delta\frac{1}{r_\eps} \ ,
\eeqan
i.e.\
 \beqa\label{4.16}
\nablav^{n+1}\vert\vert\frac{\msft^{(n)}}{r_\eps}=\nablav^{n+1}\vert\vert\frac{\mct^{(n)}}{r_\eps}
-4\pi\frac{n(n-1}{2}\,\widetilde{\latens}^{(n-2)}\vert\vert\nablav^{n-1}\delta_\eps(\rvec)\ .
\eeqa
 The last term from equation \eref{4.13} can be written as
 \beqa\label{4.17}
\fl\;\;\;\;\;\;\;\nablav^{n-1}\vert\vert\Delta\frac{\msft^{(n)}}{r_\eps}=-4\pi\,\evec_i\,\d_{i_1}\dots\d_{i_{n-1}}\,\delta_\eps(\rvec)\,\msf_{i_1\dots i_{n-1}\,i}=-4\pi\,\nablav^{n-1}\delta_\eps(\rvec)\vert\vert\msft^{(n)}\ .
\eeqa
It remains to separate the singularity of interest in the first term from the right-hand side  of equation \eref{4.16}. 
In this case, we can employ equation \eref{4.11} with the substitution $\pct\to\mct$ obtaining
\beqan
\fl\;\;\;\;\;\left(\nablav^{n+1}\vert\vert\frac{\msft^{(n)}}{r_\eps}\right)_{(0)}=-4\pi\,\frac{n}{2n+1}\mct^{(n)}\vert\vert\nablav^{n-1}\delta(\rvec)-4\pi\frac{n(n-1}{2}\,\widetilde{\latens}^{(n-2)}\vert\vert\nablav^{n-1}\delta(\rvec)\ .
\eeqan
With this last result and with equations \eref{4.13} and \eref{4.17} we can write
 \beqa\label{4.18}
\fl\left(\Bvec^{(n)}(\rvec)\right)_{(0)}=&\frac{(-1)^n\mu_0}{n!}\left[\frac{n}{2n+1}\mct^{(n)}\vert\vert\nablav^{n-1}\delta_\eps(\rvec)+\frac{n(n-1}{2}\,\widetilde{\latens}^{(n-2)}\vert\vert\nablav^{n-1}\delta(\rvec)\right.\nonumber\\
&-\left.\nablav^{n-1}\delta(\rvec)\vert\vert\msft^{(n)}\right]\ .
\eeqa
With this equation, the  separation of the $\delta-$singularities for the magnetic field can be considered finished.

\section{Concluding remarks}

Along this article, we presented an alternative and, we believe, simpler procedure to determine the $\delta-$singularities of the static electromagnetic field.  The method is based on a generalization of an identity involving the regularized derivative of $1/r$ and the regularized function $\delta_\eps(\rvec)$. Section 2 presented the discussion for the dipole, quadrupole and octopole of the static electric field and section 3 treated the magnetic case. Section 4 gave the generalization of the results for arbitrary $n$ multipoles. \\
All the results from a recent paper \cite{cvrz} are recovered in the static case. These results are obtained in Ref.\ \cite{cvrz} by a generalization of the procedure employed in Ref.\ \cite{Frahm}. The method presented in the current paper is more direct and avoids the discussions related to the regularization procedures.


\end{document}